\documentclass[showpacs,twocolumn,preprintnumbers,amsmath,amssymb]{revtex4}
\usepackage{graphicx}
\usepackage{dcolumn}
\usepackage{bm}
\usepackage{epsfig}
\begin{document}

\preprint{}

\title{Evolution of the Spiral Waves in Excitable System}

\author{Ji-Rong Ren}\thanks{Email: renjr@lzu.edu.cn.}
\author {Tao Zhu}
\thanks{Corresponding author. Email :zhut05@lzu.cn.}
\author {Shu-Fan Mo}\thanks{Email: meshf07@lzu.cn.}

\affiliation{Institute of Theoretical Physics, Lanzhou University,
Lanzhou 730000, P. R. China}

\date{\today}

\begin{abstract}
Spiral wave, whose rotation center can be regarded as a point
defect, widely exists in various two dimensional excitable systems.
In this paper, by making use of \emph{Duan's topological current
theory}, we obtain the charge density of spiral waves and the
topological inner structure of its topological charge. The evolution
of spiral wave is also studied from the topological properties of a
two-dimensional vector field. The spiral waves are found generating
or annihilating at the limit points and encountering, splitting, or
merging at the bifurcation points of the two-dimensional vector
field. Some applications of our theory are also discussed.
\end{abstract}

\pacs{03. 65. Vf, 02. 10. Kn, 82. 40. Ck}


\maketitle

\section{Introduction}
Rotating of spiral waves is a fascinating example of
self-organization in various two dimensional excitable systems
including Belousov-Zhabotinskii reaction\cite{BZ1,BZ2,BZ3} , cardiac
muscle\cite{cardiac muscle}, and CO oxidation on platinum
surfaces\cite{CO}. Recently, much effort has been put into the
development of the analytic description of excitable system, and the
spiral waves have drawn great interest and have been studied
intensively in many
ways\cite{arm1,arm2,spiral1,spiral2,zhang1,zhang2}.

The rotation centers of the spiral wave can be regarded as a point
defect and defined in terms of a phase singularity. The control of
these singular points is a very complex problem in all
two-dimensional excitable system. The dynamics of spiral wave are
determined not only by properties of the excitable system but also
by the topological characteristic number of the spiral wave. The
topological charge, which describes the strength of spiral wave
phase singularity, is a basic topological number of spiral
wave\cite{arm1,zhang,top1,top2,win,top4,top5,topologicalcharge1,topologicalcharge2,topologicalcharge3,
topologicalcharge4,topologicalcharge5,topologicalcharge6,topologicalcharge7,topologicalcharge8}.
It plays an essential role during the processes of creating or
annihilating new spiral waves. These processes can be regarded as
topological phenomena in excitable system. The mathematical
description of these spiral waves relates to the notion of phase
which in turn allows one to characterize spiral waves by an index.
The phase field $\theta(r,t)$ of spiral waves is a continuously
differentiable function except at the singular
points\cite{arm1,top1,topologicalcharge5,topologicalcharge7}. The
sites of spiral wave cores are just the phase singular points. An
index, which counts the winding number of phase field around the
singular point, is just the topological charge of the spiral wave.
From this description, a number of topological arguments have been
applied to help us to understand the topological properties of
spiral wave and some important topological constrains on behaviors
of spiral wave have been
investigated\cite{arm1,zhang,top1,top2,win,top4,top5,topologicalcharge1,topologicalcharge2,topologicalcharge3,
topologicalcharge4,topologicalcharge5,topologicalcharge6,topologicalcharge7,topologicalcharge8}.
However, the traditional topological theory of spiral wave is not so
powerful at directly computing the singularities of phase fields,
but only considers a closed path around the studied region, and uses
the single-value condition to indirectly analyze the existence of
spiral wave in the region. Therefore, it is very important to
develop a new method which can directly deduce these spiral waves in
a precise topological form, and can reveal the inner structures and
the branch processes (generate, annihilate, encounter, split, or
merge) of the spiral waves.

In this paper, by making use of \emph{Duan's topological current
theory}\cite{duan1,duan6,duan2,duan5,duan3}, we study the inner
structure of spiral wave in details. It is showed that the center of
spiral wave is not only classified by topological charge, but also
classified by Hopf index, Brouwer degree in more detail in geometry.
We also investigate the evolution of spiral wave. And one sees that
the spiral waves generate or annihilate at the limit points and
encounter, split, or merge at the bifurcation points of the spiral
wave field. Furthermore, we summarize the topological constrains on
the behaviors of spiral waves during the above processes. Some
possible applications of our topological theory of spiral wave are
also discussed.

\section{The Topological Charge Densities of Spiral waves}
Excitable system is typically described in terms of
reaction-diffusion models. We chose to work with a general
two-variable reaction-diffusion system which mathematical
description in terms of a nonlinear partial differential equation.
This equation is written as
\begin{eqnarray}
\partial_t u=F_1(u,v)+D_u\nabla^2u,\nonumber\\
\partial_t v=F_2(u,v)+D_v\nabla^2v.\label{rd}
\end{eqnarray}
Here $u$ and $v$ represent the concentrations of the reagents;
$F_1(u,v)$ and $F_2(u,v)$ are the reaction functions. The diffusion
coefficients of these reagents are $D_u$ and $D_v$. Following the
description in Ref.\cite{zhang}, we define a complex function
$Z=\phi^1+i\phi^2$, where $\phi^1=u-u^*$ and $\phi^2=v-v^*$. Here
$u^*$ and $v^*$ are the concentrations of the core of spiral wave.

As pointed out in Ref.\cite{zhang,win}, the sites of the spiral
waves are just the isolated zero points of the complex function
$Z=\phi^1+i\phi^2$, i.e., the spiral wave field in two-dimensional
surface, and in general, it is called the spiral core. The phase
field of spiral wave is defined as the argument of the complex
function $Z$, i.e., $\theta(r,t)=\arg(Z)=\arctan(\phi^1/\phi^2)$. It
is easy to see that the zero points of $Z$ are just the phase
singular points of phase field $\theta(r,t)$. By using of
topological viewpoints, Liu et al.\cite{defect} derived a
topological expression of the charge density, which is written as
\begin{eqnarray}
\rho(x,y,t)=\sum^{N}_{l=1}W_{l}\delta^2[r-r_{l}(t)],\label{char}
\end{eqnarray}
where $W_l$ is the topological charge of the $l$-th spiral wave.
This expression also reveals the topological structure of the spiral
wave. In our topological theory of spiral wave, the topological
structure of spiral wave will play an essential role. In order to
make the background of this paper clear, in this section we rewrite
the charge density (\ref{char}) of spiral wave as a topological
current.

Now we begin to derive the topological current form of the charge
density of the spiral core. We know that the complex function
$Z=\phi^1+i\phi^2$ can be regarded as the complex representation of
a two-dimensional vector field $\vec{Z}=(\phi^1,\phi^2)$. Let us
define the unit vector: $n^a=\frac{\phi^a}{\|\phi\|} (a=1,2;
\|\phi\|^2=\phi^a\phi^a=Z^*Z)$. It is easy to see that the zeros of
$Z$ are just the singularities of $\vec{n}$. Using this unit vector
$\vec{n}$, we define a two dimensional topological current
\begin{equation}
J^{i}=\frac{1}{2\pi}\epsilon^{ijk}\epsilon_{ab}\partial_{j} n^a
\partial_k n^b,~~~~i,j,k=0,1,2.\label{topcu}
\end{equation}
Applying \emph{Duan's topological current theory}, one can obtain
\begin{equation}
J^{i}=\delta^2(\vec{\phi})D^{i}(\frac{\phi}{x}),\label{delt}
\end{equation}
where the Jacobian $D^i(\frac{\phi}{x})$ is defined as
\begin{equation}
D^{i}(\frac{\phi}{x})=\frac{1}{2}\epsilon^{ijk}\epsilon_{ab}\partial_{j}
\phi^a \partial_{k} \phi^b.
\end{equation}
The delta function expression (\ref{delt}) of the topological
current $J^i$ tell us that it does not vanish only when the spiral
waves exist,i.e.,$Z=0$. The sites of the spiral core determine the
nonzero solutions of $J^i$. The implicit function theorem \cite{imp}
shows that under the regular condition
\begin{eqnarray}
D^0(\frac{\phi}{x})\neq 0,
\end{eqnarray}
the general solutions of
\begin{eqnarray}
\phi^{a}(x^{0},x^{1}.x^{2})=0,~~~~~~a=1,2,\label{aa}
\end{eqnarray}
can be expressed as
\begin{eqnarray}
&&\vec{x}=\vec{z_{l}}(t), ~~~~~~l=1,2,\cdots,N,\nonumber\\
&&~~~~~~~x^0=t.
\end{eqnarray}
From Eq.(\ref{aa}), it is easy to prove that
\begin{eqnarray}
\left.D^{i}(\frac{\phi}{x})\right|_{z_l}=\left.D(\frac{\phi}{x})\right|_{z_l}\frac{dx^{i}}{dt}.
\end{eqnarray}
According to the $\delta$-function theory \cite{deltf} and
\emph{Duan's topological current theory}, one can prove that
\begin{eqnarray}
J^{i}=\sum^{N}_{l=1}\beta_{l}\eta_{l}\delta^2(\vec{x}-\vec{z_{l}})\left.\frac{dx^{i}}{dt}\right|_{z_{l}},
\end{eqnarray}
in which the positive integer $\beta_{l}$ is the Hopf index and
$\eta_{l}=sgn(D(\frac{\phi}{x})\big|_{z_{l}})=\pm1$ is the Brouwer
degree. Now the density of spiral wave are expressed in terms of the
complex function $Z$:
\begin{eqnarray}
\rho=J^0&=&\frac{1}{2\pi}\epsilon^{0jk}\epsilon_{ab}\partial_{j} n^a
\partial_k
n^b\nonumber\\&=&\sum^{N}_{l=1}\beta_{l}\eta_{l}\delta^2(\vec{x}-\vec{z_{l}}).\label{rot}
\end{eqnarray}
This is just the charge density of spiral in Ref.\cite{zhang}.
Therefore, the total charge of the system given can be rewritten as
\begin{eqnarray}
Q=\int\rho ds^2=\sum^{N}_{l=1}W_l=\sum^{N}_{l=1}\beta_{l}\eta_{l},
\end{eqnarray}
where $W_l$ is just the winding number of $\vec{Z}$ around $z_l$,
the above expression reveals distinctly that the topological charge
of spiral wave is not only the winding number, but also expressed by
the Hopf indices and Brouwer degrees. For multi-armed spiral wave,
the Hopf indices $\beta_{l}$ characterizes the numbers of arms in a
spiral wave ($\beta_{l}=1$, a one-armed spiral wave; $\beta_{l}=2$,
a two-armed spiral wave). And the Brouwer degree $\eta_{l}$, whose
sign is very important, characterizes the direction of rotation of
spiral wave ($\eta_{l}=+1$, the spiral wave rotating clockwise;
$\eta_{l}=-1$, he spiral wave rotating anti-clockwise). It is seen
that our topological theory provides a more detailed description of
the inner structures of spiral wave. The topological inner structure
showed in Eq. (\ref{rot}) is more essential than that in Eq.
(\ref{char}), this is just the advantage of our topological
description of the spiral wave.

According to Eq.(\ref{topcu}), it is easy to see that the
topological current $J^i$ is identically conserved,
\begin{eqnarray}
\partial_iJ^{i}=0.
\end{eqnarray}
This equation implies the conservation of the topological charge of
spiral wave:
\begin{eqnarray}
\partial_t\rho+\nabla\cdot \vec{J}=0,
\end{eqnarray}
which is only the topological property of the complex function $Z$.
The conservation of the total topological charge $Q$ is a very
stronger topological constraint on spiral wave, many important
properties of spiral waves due to this topological rule. In our
following sections, we discuss the generating, annihilating at the
limit points and encountering, splitting, merging at the bifurcation
points of complex function $Z$, and it shows that the conservation
of the total topological charge $Q$ is also valid in these
processes.

\section{The generation and annihilation of spiral waves}
As investigated before, the zeros of the complex function $Z$ play
an important role in describing the topological structures of spiral
waves. Now we begin discussing the properties of the zero points, in
other words, the properties of the solutions of Eq. (\ref{aa}). As
we knew before, if the Jacobian
\begin{equation}
D^0(\frac{\phi}{x}) \neq 0,\label{reu}
\end{equation}
we will have the isolated zeros of the vector field $\vec{Z}$. The
isolated solutions are called regular points. However, when the
condition (\ref{reu}) fails, the usual implicit function theorem
\cite{imp} is of no use. The above discussion will change in some
way and will lead to the branch process. We denote one of the zero
points as $(t^*,\vec{z_i})$. If the Jacobian
\begin{equation}
\left.D^1(\frac{\phi}{x})\right|_{(t^*,\vec{z_l})}\neq 0,
\end{equation}
we can use the Jacobian $D^1(\frac{\phi}{x})$ instead of
$D^0(\frac{\phi}{x})$ for the purpose of using the implicit function
theorem. Then we have a unique solution of Eq.(\ref{aa}) in the
neighborhood of the limit point $(t^*,\vec{z}_l)$
\begin{equation}
t=t(x^1),~~~~~x^2=x^2(x^1),
\end{equation}
with $t^*=t(z^1_l)$. We call the critical points $(t^*,\vec{z_l})$
the limit points. In the present case, we know that
\begin{equation}
\left.\frac{dx^1}{dt}\right|_{(t^*,\vec{z}_l)}=\left.\frac{D^1(\frac{\phi}{x})}{D(\frac{\phi}{x})}\right|_{(t^*,\vec{z}_l)}=\infty
\label{wuqiong}
\end{equation}
i.e.,
\begin{equation}
\left.\frac{dt}{dx^1}\right|_{(t^*,\vec{z}_l)}=0
\end{equation}
Then the Taylor expansion of $t=t(x^1)$ at the limit point
$(t^*,\vec{z}_l)$ is
\begin{equation}
t-t^*=\frac{1}{2}\left.\frac{d^2
t}{(dx^1)^2}\right|_{(t^*,\vec{z}_l)}(x^1-z^1_l)^2,\label{cha}
\end{equation}
which is a parabola in $x^1-t$ plane. From Eq.(\ref{cha}) we can
obtain two solutions $x^1_1(t)$ and $x^1_2(t)$, which give two
branch solutions (world lines of the spiral waves). If
\begin{equation}
\left.\frac{d^2 t}{(dx^1)^2}\right|_{(t^*,\vec{z}_l)}>0,
\end{equation}
We have the branch solutions for $t>t^*$ [see Fig.1(a)]; otherwise,
we have the branch solutions for $t<t^*$ [see Fig.1(b)]. These two
cases are related to the origin and annihilation of the spiral
waves.
\begin{figure}[h]
\begin{center}
\begin{tabular}{p{5cm}p{5cm}}
(a)\psfig{figure=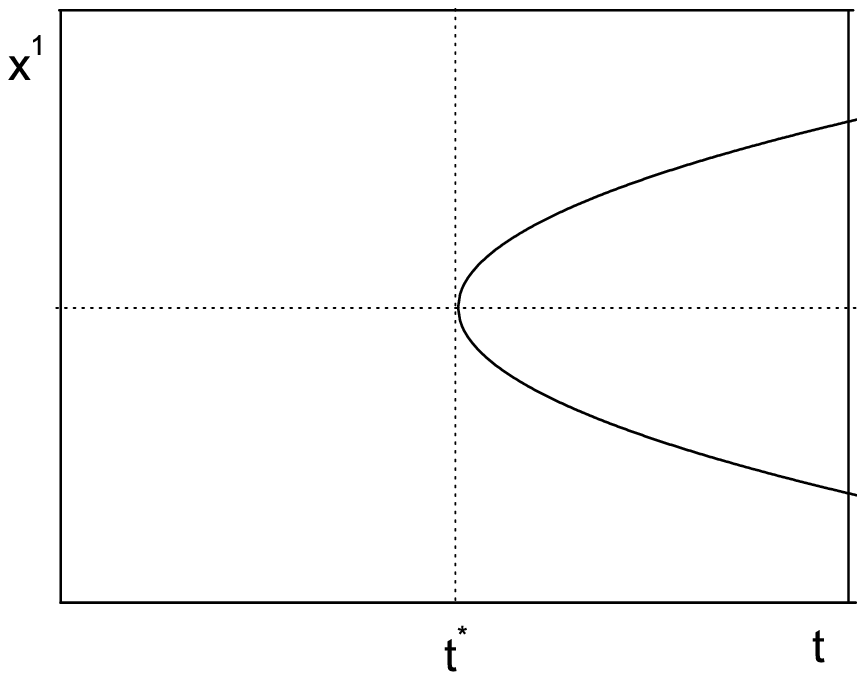,height=5cm,width=5cm}\\
(b)\psfig{figure=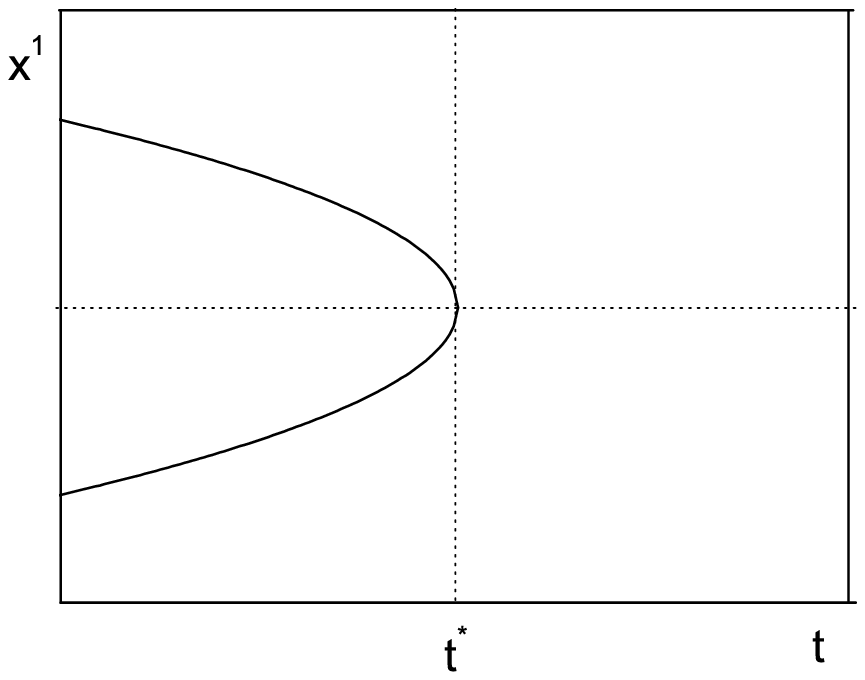,height=5cm,width=5cm}\\
\end{tabular}
\end{center}
\caption{Projecting the world lines of spiral waves onto $(x^1-t)$
plane. (a) The branch solutions for Eq.(\ref{cha}) when
$d^2t/(dx^1)^2|_{(t^*,\vec{z}_l)}>0$, i.e., a pair of spiral waves
with opposite charges generate at the limit point, i.e., the origin
of spiral waves. (b) The branch solutions for Eq.(\ref{cha}) when
$d^2t/(dx^1)^2|_{(t^*,\vec{z}_l)}<0$, i.e., a pair of spiral waves
with opposite charges annihilate at the limit point.} \label{fig1}
\end{figure}

One of the results of Eq.(\ref{wuqiong}), that the velocity of the
spiral waves are infinite when they are annihilating, agrees with
the fact obtained by Bray \cite{bray} who has a scaling argument
associated with the point defects final annihilation which leases to
a large velocity tail. From Eq.(\ref{wuqiong}), we also obtain a new
result that the velocity field of spiral waves is infinite when they
are generating, which is gained only from the topology of the
complex function $Z$.

Since topological current is identically conserved, the topological
charge of these two generated or annihilated spiral pair must be
opposite at the limit point, i.e.,
\begin{equation}
\beta_{l_1}\eta_{l_1}=-\beta_{l_2}\eta_{l_2}\label{cc1}
\end{equation}
which shows that $\beta_{l_1}=\beta_{l_2}$ and
$\eta_{l_1}=-\eta_{l_2}$. One can see that the fact the Brouwer
degree $\eta$ is indefinite at the limit points implies that it can
change discontinuously at limit points along the world lines of the
spiral vortices (from $\pm1$ to $\mp1$). It is easy to see from Fig.
1:when $x^1>z^1_{l},\eta_{l_1}=\pm1$; when
$x^1<z^1_{l},\eta_{l_1}=\mp1$.

For a limit point it is required that
$D^1(\frac{\phi}{x})\big|_{(t^*,\vec{z}_l)}\neq 0$. As to a
bifurcation point \cite{bif}, it must satisfy a more complex
condition. This case will be discussed in the following section.

\section{Bifurcation of the velocity field of the phase singularity for spiral wave}

In this section we have the restrictions of Eq. (\ref{aa}) at the
bifurcation points $(t^{*},\vec{z_{l}})$,
\begin{equation}
\left.D(\frac{\phi}{x})\right|_{z_{l}}=0,~~~~\left.D^1(\frac{\phi}{x})\right|_{z_{l}}=0,\label{restrict}
\end{equation}
which leads to an important fact that the function relationship
between $t$ and  $x^1$ is not unique in the neighborhood of the
bifurcation point $(t^{*}, \vec{z_{l}})$. It is easy to see that
\begin{equation}
V^1=\frac{dx^1}{dt}=\left.\frac{D^1(\frac{\phi}{x})}{D(\frac{\phi}{x})}\right|_{z_{l}},\label{velocity}
\end{equation}
which under Eq.(\ref{restrict}) directly shows that the direction of
the integral curve of Eq.(\ref{velocity}) is indefinite at $(t^{*},
\vec{z_{l}})$, i.e., the velocity field of the phase singularity for
spirals is indefinite at $(t^{*},\vec{z_{l}})$. That is why the very
point $(t^{*}, \vec{z_{l}})$ is called a bifurcation point.

Assume that the bifurcation point $(t^{*}, \vec{z_{l}})$ has been
found from Eqs.(\ref{aa}) and (\ref{restrict}). We know that, at the
bifurcation point $(t^{*}, \vec{z_{l}})$, the rank of the Jacobian
matrix $[\frac{\partial\phi}{\partial x}]$ is 1. In addition,
according to the $\phi-$mapping theory, the Taylor expansion of the
solution of $\phi^1$ and $\phi^2$ in the neighborhood of the
bifurcation point can generally be denoted as
\begin{equation}
A(x^1-z^1_l)^2+2B(x^2-z^2_l)(t-t^*)+(t-t^*)^2=0,
\end{equation}
which leads to
\begin{equation}
A(\frac{dx^1}{dt})^2+2B\frac{dx^1}{dt}+C=0\label{shun}
\end{equation}
and
\begin{equation}
C(\frac{dt}{dx^1})^2+2B\frac{dt}{dx^1}+A=0,\label{dao}
\end{equation}
where $A$, $B$, and $C$ are three constants.The solution of
Eq.(\ref{shun}) or Eq.(\ref{dao}) give different directions of the
branch curves (world lines of the spiral waves) at the bifurcation
point. There are four kinds of important cases, which will show the
physical meanings of the bifurcation points.

Case 1($A\neq 0$). For $\Delta =4(B^2-AC)>0$, we get two different
directions of the velocity field of the phase singularity for spiral
waves
\begin{equation}
\left.\frac{dx^1}{dt}\right|_{1,2}=\frac{-B\pm
\sqrt{B^2-AC}}{A}\label{case1}
\end{equation}
which is shown in Fig.2. It is the intersection of two phase
singularity for spirals, which means that two phase singularity for
spirals meet and then depart from each other at the bifurcation
point.

\begin{figure}[h]
\begin{center}
\begin{tabular}{p{5cm}p{5cm}}
\psfig{figure=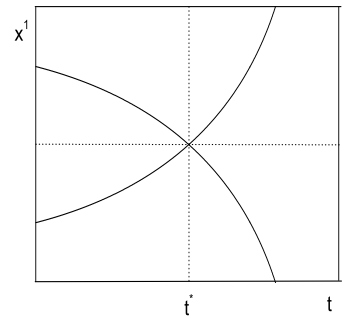,height=5cm,width=5cm}

\end{tabular}
\end{center}
\caption{Projecting the world lines of spiral waves onto $(x^1-t)$
plane. Two spiral waves meet and then depart at the bifurcation
point.}\label{fig2}
\end{figure}

Case 2($A\neq 0$). For $\Delta =4(B^2-AC)=0$, the direction of the
velocity field of the phase singularity is only one
\begin{equation}
\left.\frac{dx^1}{dt}\right|_{1,2}=\frac{-B}{A}\label{case2}
\end{equation}
which includes three important situations. (a) One world line
resolves into two world lines, i.e., one spiral wave splits into a
spiral pair at the bifurcation point [see Fig.3(a)]. (b) Two world
lines merge into one world line, i.e., a spiral pair merge into one
spiral at the bifurcation point[see Fig.3(b)]. (c) Two world lines
tangentially contact, i.e., a spiral pair tangentially encounter at
the bifurcation point[see Fig.3(c)].

\begin{figure}[h]
\begin{center}
\begin{tabular}{p{5cm}p{5cm}p{5cm}}
(a)\psfig{figure=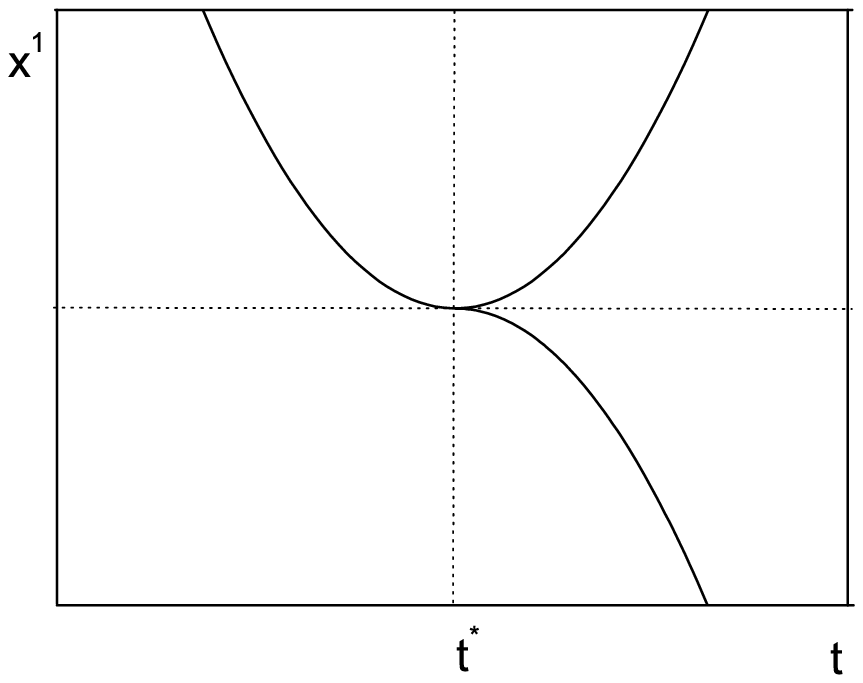,height=5cm,width=5cm} \\
(b)\psfig{figure=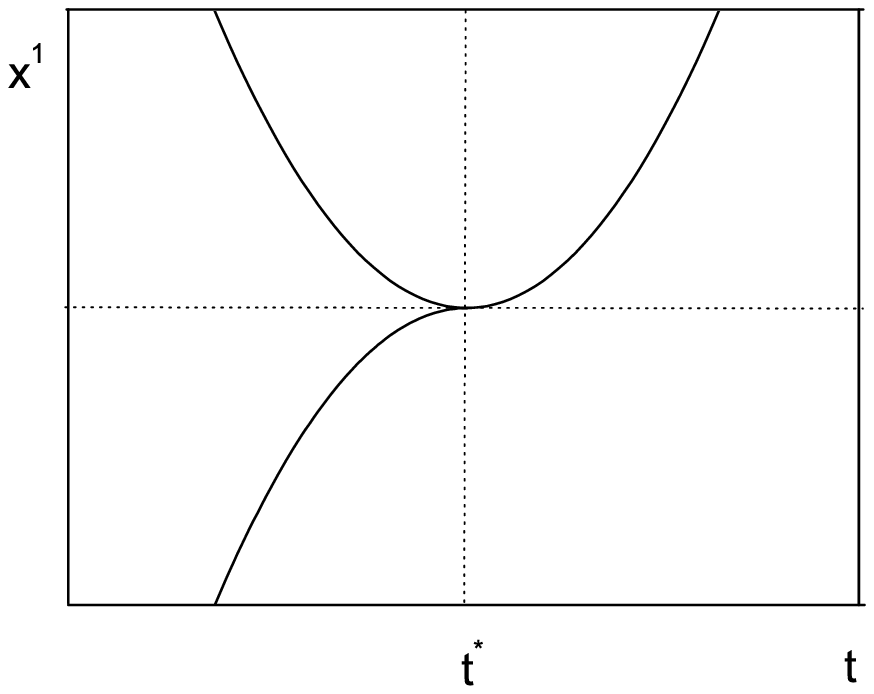,height=5cm,width=5cm} \\
(c)\psfig{figure=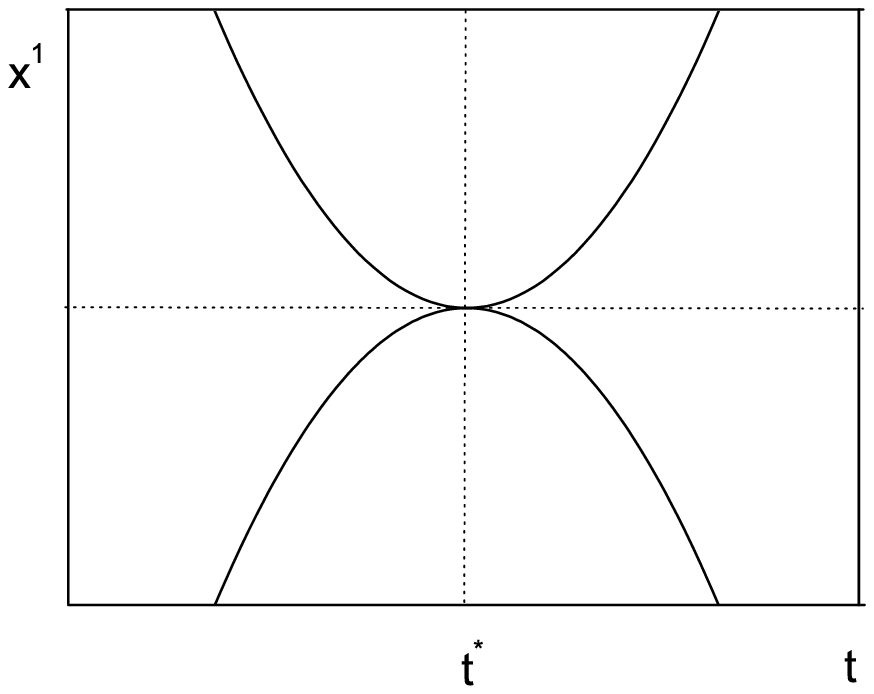,height=5cm,width=5cm}
\end{tabular}
\end{center}
 \caption{ (a) One spiral wave splits into two spiral wave at the bifurcation point. (b) Two
spiral waves merge into one spiral wave at the bifurcation point.
(c) Two world line of spiral waves tangentially intersect, i.e., two
spiral waves tangentially encounter at the bifurcation point.}
\label{fig3}
\end{figure}

Case 3($A=0, C\neq 0$). For $\Delta=4(B^2-AC)=0$, we have
\begin{equation}
\left.\frac{dt}{dx^1}\right|_{1,2}=\frac{-B\pm\sqrt{B^2-AC}}{C}=0,~~~-\frac{2B}{C}\label{case3}
\end{equation}
There are two important cases: (a) Three world lines merge into one
world line, i.e., three spirals merge into a spiral at the
bifurcation point[see Fig.4(a)]. (b) One world line resolves into
three world lines, i.e., a spiral splits into three spirals at the
bifurcation point[see Fig.4(b)].

Case 4($A=C=0$). Equation(\ref{shun}) and Eq(\ref{dao}) give
respectively
\begin{equation}
\frac{dx^1}{dt}=0,~~~\frac{dt}{dx^1}=0.\label{case4}
\end{equation}
This case is obvious, see Fig.5, and is similar to Case 3.

The above solution reveal the evolution of the spirals. Besides the
encountering of the spiral waves, i.e., a spiral pair encounter and
then depart at the bifurcation point along different branch cures
[see Fig.2 and Fig.3(c)], it also includes splitting and merging of
spirals. When a multi-charged spiral moves through the bifurcation
point, it may split into several spirals along different branch
curves [see Fig.3(a), Fig.4(b), Fig.5(b)]. On the contrary, several
spirals can merge into a spiral at the bifurcation point [see
Fig.3(b) and Fig.4(a)].

\begin{figure}[h]
\begin{center}
\begin{tabular}{p{5cm}p{5cm}}
(a)\psfig{figure=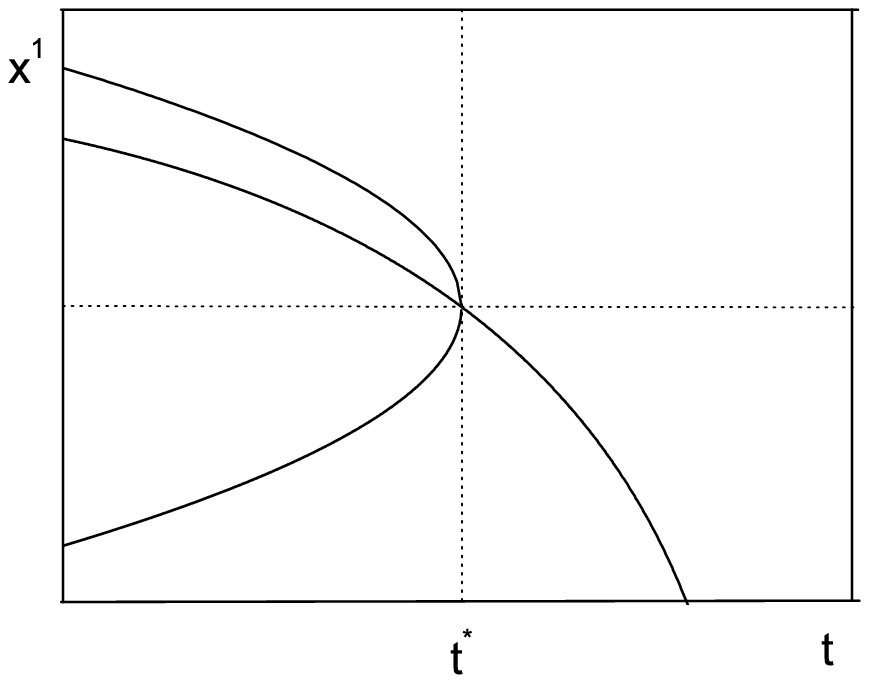,height=5cm,width=5cm} \\
(b)\psfig{figure=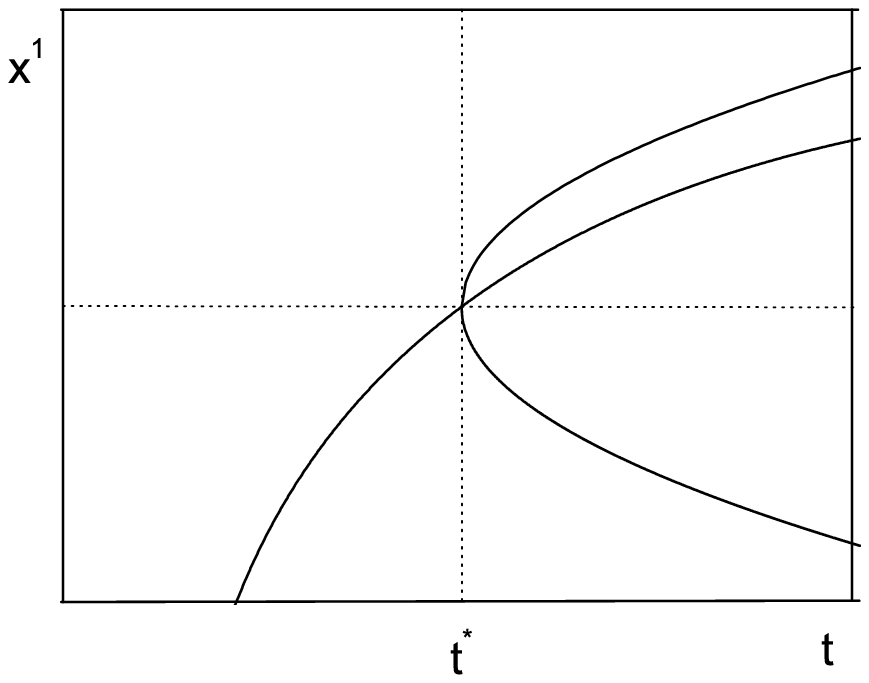,height=5cm,width=5cm}
\end{tabular}
\end{center}
\caption{ Two important cases of Eq.(\ref{case3}). (a) Three spiral
waves merge into one at the bifurcation point. (b) One spiral wave
splits into three spiral waves at the bifurcation point.}
\label{fig4}
\end{figure}

\begin{figure}[h]
\begin{center}
\begin{tabular}{p{5cm}p{5cm}}
(a)\psfig{figure=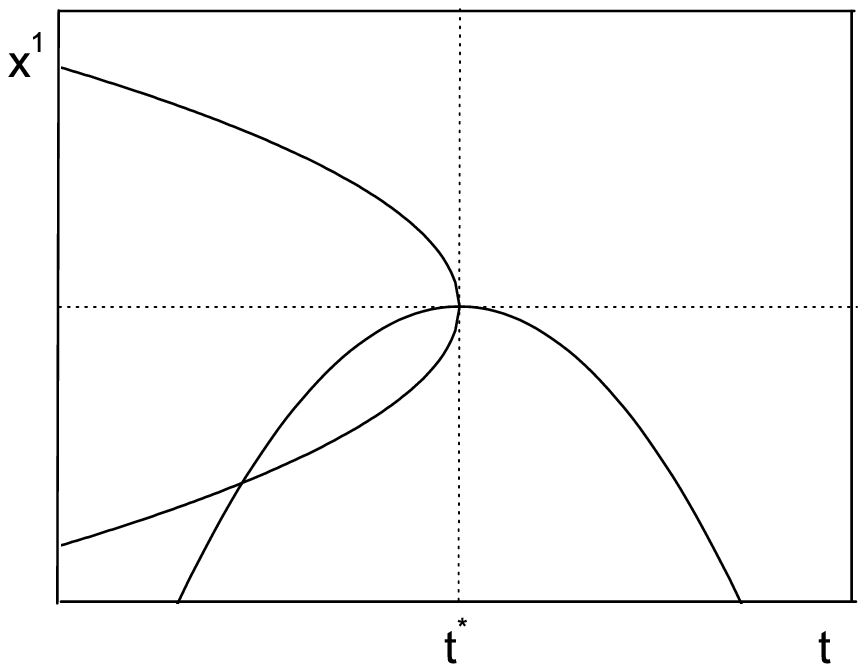,height=5cm,width=5cm} \\
(b)\psfig{figure=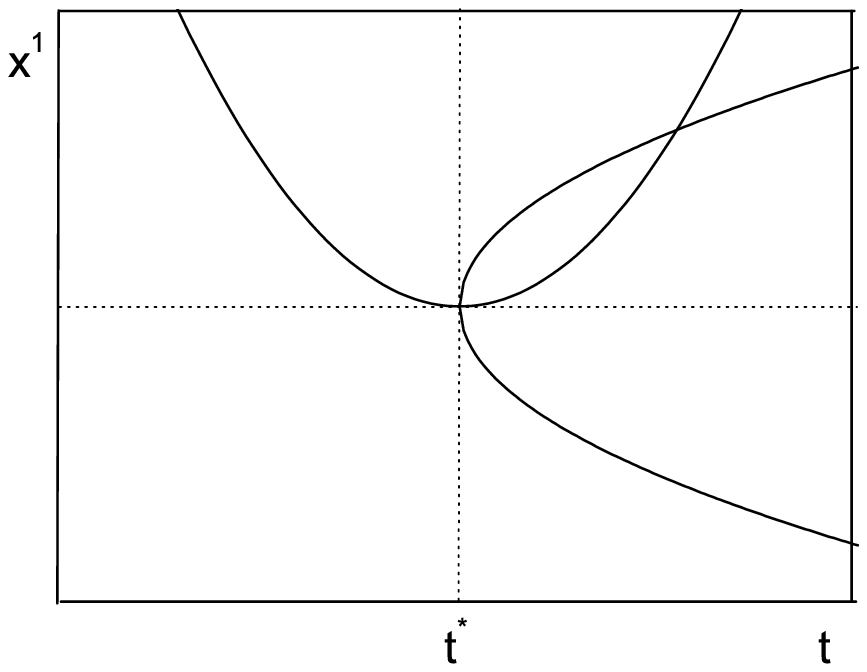,height=5cm,width=5cm}
\end{tabular}
\end{center}
\caption{ Two world lines intersect normally at the bifurcation
point. This case is similar to FIG. 4. (a) Three spiral waves merge
into one at the bifurcation point. (b) One spiral wave splits into
three spiral waves at the bifurcation point.} \label{fig5}
\end{figure}

The identical conversation of the topological charge shows the sum
of the topological charge of these final spirals must be equal to
that of the original spirals at the bifurcation point, i.e.,
\begin{equation}
\sum_{i}\beta_{l_{i}}\eta_{l_{i}}=\sum_{f}\beta_{l_{f}}\eta_{l_{f}}\label{cc2}
\end{equation}
for fixed $l$. Furthermore, from the above studies, we see that the
generation, annihilation, and bifurcation of spirals are not
gradually changed, but suddenly changed at the critical points.

\section{Application and Discussion}
Our conclusions can be summarized as follows: First, by making use
of \emph{Duan's topological current theory}, we obtain the charge
density of spiral waves and the topological inner structure of its
topological charge, the spiral waves can be classified not only by
their topological charges, but especially importantly by their Hopf
indices and Brouwer degrees. This is more general than usually
considered and will be helpful as a complement of the current
description of spiral wave only by the winding numbers in topology.
Second, the evolution of spiral wave is also studied from the
topological properties of a two-dimensional vector field $\vec{Z}$.
We find that there exist crucial cases of branch processes in the
evolution of the spiral wave when $D(\frac{\phi}{x})\neq 0$, i.e.,
$\eta_l$ is indefinite. This means that the spiral waves are
generate or annihilate at the limit points and encounter, split, or
merge at the bifurcation points of the two-dimensional vector field
$\vec{Z}$, which shows that the spiral wave system is unstable at
these branch points. Third, we found the result that the velocity of
spiral core is infinite when they are annihilating or generating,
which is obtained only from the topological properties of the
two-dimensional vector field $\vec{Z}$. Forth, we must pointed out
that there exist two restrictions of the evolution of spiral waves.
One restriction is the conservation of the topological charge of the
spiral waves during the branch process [see Eqs. (\ref{cc1}) and
(\ref{cc2})], the other restriction is that the number of different
directions of the world lines of spiral waves is at most $4$ at the
bifurcation points [see Eqs. (\ref{shun}) and (\ref{dao})]. The
first restriction is already known, but the second is pointed out
here for the first time to our knowledge. We hope that it can be
verified in the future. Finally, we would like to point out that all
the results in this paper have been obtained only from the viewpoint
of topology without using any particular models or hypothesis and do
not require the knowledge of the actual solution of the
reaction-diffusion equation.

In this paper, we give a rigorous and general topological
investigation of spiral wave. This work can be applied in several
ways. First, topology and geometry of two-dimensional manifold in
which the spiral lived can affect the dynamics of spiral
wave\cite{spiral1,spiral2}, therefore, it is very interesting to
extend our work to a curved surface which have complex topology and
geometry, and study the interaction between the topology of spiral
wave and the topology of curved surface. Second, our theory can be
easily extend to the three dimensional scroll wave, which is just a
three dimensional analog of spiral wave, in this situation, a more
complex topology related to knots must be considered\cite{duan2}.
Third, our work provide a more detailed description of spiral wave,
the detailed information of spiral can applied to the study of
dynamics of spiral wave. These issues are very worthwhile to
consider and will be investigated in our further works.
\begin{acknowledgments}
Thanks to the works of Dr. X. H. Zhang and Dr. B. H. Gong in drawing
the figures in this paper. This work was supported by the National
Natural Science Foundation of China and the Cuiying Programme of
Lanzhou University.
\end{acknowledgments}


\begin{thebibliography}{99} \addcontentsline{toc}{section}{Bibliography}
\bibitem{BZ1}A. N. Zaikin and A. M. Zhabotinsky, Nature \textbf{225}, 535 (1970).
\bibitem{BZ2}A.T. Winfree, Science \textbf{175}, 634 (1972).
\bibitem{BZ3}Th. Plesser, S. C. Miiller and B. Hess, J. Phys. Chem. \textbf{94}, 7501 (1990).
\bibitem{cardiac muscle}J. M. Davidenko, A. V. Pertsov, R. Salomonsz, W. Baxter
and J. Jalife, Nature \textbf{355}, 349 (1992).
\bibitem{CO}S. Jakubith, H. H. Rotermund, W. Engel, A. von Oertzen
and G. Ertl, Phys. Rev. Lett. \textbf{65}, 3013 (1990).
\bibitem{arm1}K. I. Agladze and V. I. Krinsky, Nature \textbf{296}, 424
(1982); V. I. Krinsky and K. I. Agladze, Dokl. Akad. Nauk SSSR
\textbf{263}, 335 (1982).
\bibitem{arm2}B. Vasiev, F. Siegert, and C. Weijer, Phys. Rev. Lett. \textbf{78}, 2489
(1997).
\bibitem{spiral1}J. Davidsen, L. Glass, and R. Kapral, Phys. Rev. E \textbf{70}, 056203
(2004).
\bibitem{spiral2}K. Rohlf, L. Glass, and R. Kapral, Chaos \textbf{16}, 037115
(2006).
\bibitem{zhang1}H. Zhang, Z. Cao, N. J. Wu, H. P. Ying, and G. Hu,
Phys. Rev. Lett. \textbf{94}, 188301 (2005).
\bibitem{zhang2}H. Zhang, B. Hu, G. Hu, and J. Xiao, J. Chem. Phys.
\textbf{121}, 7276 (2004); J. X. Cheng, H. Zhang, and Y. Q. Li, J.
Chem. Phys. \textbf{124}, 014505 (2005).
\bibitem{zhang}H. Zhang, B. Hu, B. W. Li, and Y. S. Duan, Chin.
Phys. Lett. \textbf{24}, 1618 (2007).
\bibitem{top1}A. T. Winfree, \emph{The Geometry of Biological Time} (Springer-Verlag, New York,
2001).
\bibitem{top2}L. Glass, Science \textbf{198}, 321 (1977).
\bibitem{win}A. T. Winfree and S. H. Strogatz, Physica D \textbf{8}, 35
(1983); Physica D \textbf{9}, 65 (1983); Physica D \textbf{9}, 335
(1983); Physica D \textbf{13}, 221 (1984).
\bibitem{defect}F. Liu and F. G. Mazenko, Phys. Rev. B \textbf{46},
5963 (1992).
\bibitem{top4}D. Sumners, in \emph{Graph Theory and Topology in Chemistry},
edited by R. B. King and D. Rouvray (Elsevier, Amsterdam, 1987), p.
3.
\bibitem{top5}I. Cruz-White, Ph. D. thesis, Florida State University,
2003.
\bibitem{topologicalcharge1}A. T. Winfree, Sci. Am. \textbf{230}, 82 (1974).
\bibitem{topologicalcharge2}P. J. Nandapurkar and A. T. Winfree,
Physica D \textbf{29}, 69 (1987).
\bibitem{topologicalcharge3}A. T. Winfree, \emph{When Time Breaks
Down} (Princeton University Press, 1987).
\bibitem{topologicalcharge4}A. T. Winfree, SIAM Rev. \textbf{32}, 1
(1990).
\bibitem{topologicalcharge5}R. A. Gray, A. M. Pertsov, and J. Jalife, Nature \textbf{392},
75 (1998).
\bibitem{topologicalcharge6}A. T. Winfree, Nature \textbf{371},
233 (1994); Physica D \textbf{84}, 126 (1995).
\bibitem{topologicalcharge7}Y. Kuramoto, \emph{Chemical Oscillations, Waves and Turbulence} (Springer,
Berlin, Heidelberg, 1984).
\bibitem{topologicalcharge8}A. S. Mikhailov, \emph{Foundations of Synergetics I: Distributed Active Systems}
(Springer-Verlag, New York, 1994).

\bibitem{duan1}L. B. Fu, Y. S. Duan, and H. Zhang, Phys. Rev. D \textbf{52},
045004 (2000).
\bibitem{duan6}Y. S. Duan, S. L. Zhang,  and S. S.
Feng, J. Math. Phys 35 4463 (1994).

\bibitem{duan2}J. R. Ren, T. Zhu, and Y. S. Duan,
Chin. Phys. Lett. \textbf{25}, 353 (2008) [arXiv:0712.4196].

\bibitem{duan5}S. F. Mo, J. R. Ren, and T. Zhu, J. Phys. A: Math. Theor. 41,
315214 (2008) [arXiv:0807.2784]; W. K. Qi, T. Zhu, Y. Chen, and J.
R. Ren, arXiv:0805.4661; J. R. Ren, T. Zhu, and S. F. Mo,
arXiv:0712.4198; J. R. Ren, T. Zhu, and Y. S. Duan, Commun. Theor.
Phys. 50 (2008)345-348 [arXiv:0712.4195]; J. R. Ren, T. Zhu, and Y.
S. Duan, Chin. Phys. Lett.25, 367-370 (2008) [arXiv:0707.4529].

\bibitem{duan3}Y. S. Duan, X. Liu, and L. B. Fu, Phys. Rev. D \textbf{67},
085022 (2003).
\bibitem{imp}E. Goursat, \emph{A Course in Mathematical Analysis},
translated by E. R. Hedrick (Dover, New York, 1904), Vol. I.
\bibitem{deltf}J. A. Schouton, \emph{Tensor Analysis for Physicists} (Clarendon, Oxford,
1951).
\bibitem{bray}A. J. Bray, Phys. Rev. E \textbf{55}, 5297 (1997).
\bibitem{bif}M. Kubicek and M. Marek, \emph{Computational Methods in
Bifurcation Theory and Dissipative Structures} (Springer-Verlag, New
York, 1983).
\end{thebibliography}
\end{document}